\documentclass[11pt]{article}

\usepackage[letterpaper,top=1in,bottom=1in,left=1in,right=1in]{geometry}

\usepackage[T1]{fontenc}
\usepackage[utf8]{inputenc}
\usepackage{lmodern}

\usepackage{amsmath,amssymb}
\usepackage{graphicx}

\usepackage[round,authoryear]{natbib}
\usepackage[labelfont=bf,labelsep=period]{caption}

\usepackage{titling}
\pretitle{\begin{flushleft}\LARGE\bfseries}
\posttitle{\end{flushleft}}
\preauthor{\begin{flushleft}\large}
\postauthor{\end{flushleft}}
\predate{}\postdate{}

\usepackage{titlesec}
\titlespacing*{\section}{0pt}{1.1ex plus .4ex}{0.6ex}
\titleformat{\section}{\large\bfseries}{ }{0pt}{}
\titlespacing*{\subsection}{0pt}{1.0ex plus .3ex}{0.5ex}
\titleformat{\subsection}{\normalsize\bfseries}{ }{0pt}{}

\usepackage{xcolor}

\usepackage[
  colorlinks=true,
  linkcolor=blue,   % section/eq/figure/page refs
  citecolor=blue,   % citation link targets
  urlcolor=blue,    % URLs/DOIs
  breaklinks=true
]{hyperref}
\hypersetup{pdfborder={0 0 0}} % no rectangles around links

\usepackage{doi}

\makeatletter

\let\origdoi\doi
\renewcommand{\doi}[1]{\textcolor{blue}{\origdoi{#1}}}
\makeatother

\renewcommand{\cite}{\citep}

\makeatletter
\def\NAT@nmfmt#1{\textcolor{blue}{#1}}  % author names in blue
\let\NAT@openSaved\NAT@open
\let\NAT@closeSaved\NAT@close
\renewcommand{\NAT@open}{\NAT@openSaved\begingroup\color{blue}}
\renewcommand{\NAT@close}{\endgroup\NAT@closeSaved}
\let\CiteYearOrig\citeyear
\renewcommand{\citeyear}[1]{\textcolor{blue}{\CiteYearOrig{#1}}}
\let\CiteYearParOrig\citeyearpar
\renewcommand{\citeyearpar}[1]{\textcolor{blue}{\CiteYearParOrig{#1}}}
\makeatother
\let\RefOrig\ref
\renewcommand{\ref}[1]{\textcolor{blue}{\RefOrig{#1}}}
\let\EqrefOrig\eqref
\renewcommand{\eqref}[1]{\textcolor{blue}{\EqrefOrig{#1}}}
\let\PagerefOrig\pageref
\renewcommand{\pageref}[1]{\textcolor{blue}{\PagerefOrig{#1}}}

\title{GammaPBHPlotter: A public code for calculating the complete Hawking evaporation gamma-ray spectra from primordial black holes}

\author{John Carlini$^{1}$, and Ilias Cholis$^{1}$\\[0.5em]
\small $1$ Department of Physics, Oakland University, Rochester, Michigan, 48309, USA
\\
\small %\texttt{jcarlini@oakland.edu} \quad \texttt{cholis@oakland.edu}
}

\date{} % no date

\begin{document}
\maketitle

\section{Summary}
We present \texttt{GammaPBHPlotter}, a public Python code for calculating and plotting the Hawking radiation gamma-ray spectra of primordial black holes in the mass range of $10^{14}$ to $10^{18}$ grams. This tool allows users to compute the monochromatic and mass-averaged spectra of black holes over a range of parameters. We include the primary/direct Hawking emission, the secondary emission from the decay and hadronization of unstable particles, the final state radiation, and the in-flight annihilation gamma-ray emission components.

\section{Statement of Need}

Hawking radiation \cite{1974Natur.248...30H} remains an unobserved property of black holes.
As the temperature of black holes is inversely proportional to the square of their mass, conventional stellar mass black holes  are expected to emit too little radiation to ever be detected. However, primordial black holes (PBHs) that could have formed from the collapse of primordial perturbations in the early universe can provide detectable signals \cite{Carr:2016drx}. PBHs with mass less than $10^{14}$ grams would have evaporated via Hawking radiation long before the present age of the universe. Upcoming gamma-ray telescopes such as e-ASTROGAM \cite{e-ASTROGAM:2017pxr} and AMEGO-X \cite{Caputo:2022xpx} will be sensitive enough in the MeV range to detect the Hawking spectra of PBHs lying between this lower bound and $10^{19}$ grams. 
We have developed \texttt{GammaPBHPlotter}, an open-source software to simulate the exact gamma-ray spectra produced from different PBH mass-distributions. 

\section{Hawking Spectra}
\subsection{Modeling the emission components}
The gamma-ray spectrum of a PBH within the relevant mass range consists of four primary components; direct/primary Hawking radiation, secondary radiation, final-state radiation, and in-flight annihilation.

Direct Hawking radiation accounts for all kinematically allowed elementary particles formed at the event Horizon \cite{1974Natur.248...30H}, including gamma-ray photons.
Secondary radiation originates from the decay of unstable particles and contributes significantly at lower energies. We rely on \texttt{BlackHawk} \cite{Arbey:2021mbl} to evaluate the gamma-ray primary and secondary spectral components. 
\texttt{BlackHawk} uses \texttt{PYTHIA} \cite{Sjostrand:2014zea} for the modeling of the hadronization and decay processes leading to the secondary spectra.
Final-state radiation originates from relativistic electrons and positrons and has a differential spectrum given by Eq.~\ref{eq:FSRRate},
\begin{equation}
\frac{dN_{\gamma}^{\text{FSR}}}{dE_{\gamma}} = \frac{\alpha}{2\pi} \int dE_{e^{+}} \frac{dN_{e^{+}}}{dE_{e^{+}}} \left( \frac{2}{E_{\gamma}} + \frac{E_{\gamma}}{E_{e^{+}}^2} - \frac{2}{E_{e^{+}}} \right) \left[ \ln \left( \frac{2E_{e^{+}}+(E_{e^{+}} - E_{\gamma})}{m_{e^{+}}^2} \right) - 1 \right],
\label{eq:FSRRate}
\end{equation}
where $\alpha = 137.037$ is the fine structure constant, $E_{e^{+}}$ is the kinetic energy of a given positron ($e^{+}$), $E_{\gamma}$ is the energy of the emitted photon, $m_{e^{+}} = 0.511$ MeV is the rest mass of the electron, and $\frac{dN_{e^{+}}}{dE_{e^{+}}}$ the differential spectrum of emitted electrons/positrons. 
In addition to the previously mentioned components, gamma-rays can be produced through pair-annihilation of positrons with interstellar medium electrons. This is known as in-flight annihilation and its differential spectrum is \cite{Keith:2022sow},
\begin{equation}
\frac{dN_{\gamma}^{\text{IA}}}{dE_{\gamma}} = \frac{\pi \alpha^2 n_H}{m_e} \int_{m_e}^{\infty} dE_{e^{+}} \frac{dN_{e^{+}}}{dE_{e^{+}}} \int_{E_{\min}}^{E_{e^{+}}} \frac{dE}{dE/dx} \frac{P_{E_{e^{+}} \to E}}{(E^2 - m_e^2)}
\end{equation}

\[
\times \left( -2 - \frac{(E + m_e)(m_e^2 (E + m_e) + E_{\gamma}^2 (E + 3m_e) - E_{\gamma} (E + m_e)(E + 3m_e))}{E_{\gamma}^2 (E - E_{\gamma} + m_e)^2} \right).
\]
We take $n_H = 1\, {\textrm{cm}^{-3}}$ as the density of interstellar medium hydrogen (and by extension electrons). $E_{e^{+}}$ is again the initial positron total energy, $E$ is the final positron total energy, $dE/dx$ is the rate of positron energy lost per path via the Bethe-Bloch formula \cite{Bethe1953}, $E_{\gamma}$ is the resulting photon energy from annihilation, and $P_{E_{e^{+}} \to E}$ is the probability of a particular positron of a given initial and final energy to decay. This probability matrix can be calculated as 
\cite{Keith:2022sow},
\begin{equation}
P_{E_{e^{+}} \to E} = \exp\Biggl( 
  - n_H \int_{E}^{E_{e^{+}}} \sigma_{\mathrm{ann}}(E') \,\frac{dE'}{dx}\, dE' 
\Biggr),
\end{equation}
where $\sigma_{ann}$ is the cross section of annihilation for positrons of a given energy.

\begin{figure}
    \centering
    \includegraphics[width=0.6\linewidth]{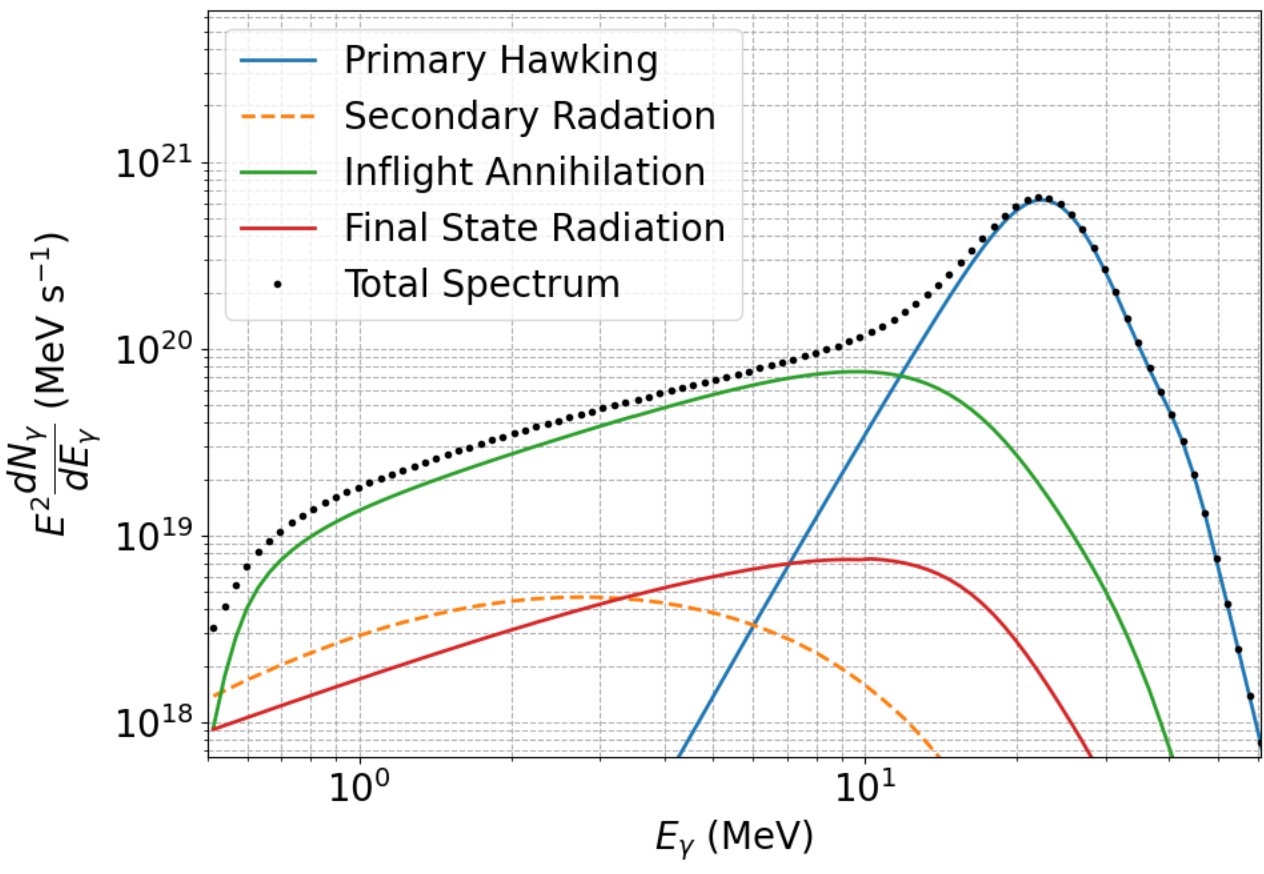}
    \caption{The total gamma-ray spectrum of a $3\times10^{15}$ grams PBH as well as its components.}
    \label{fig:PBHspectral_components}
\end{figure}

In Fig.~\ref{fig:PBHspectral_components}, we give the individual gamma-ray spectral components as well as their sum for a PBH of mass $3\times 10^{15}$ grams.

\subsection{PBH Mass Distribution}
\begin{figure}
    \centering
    \includegraphics[width=0.6\linewidth]{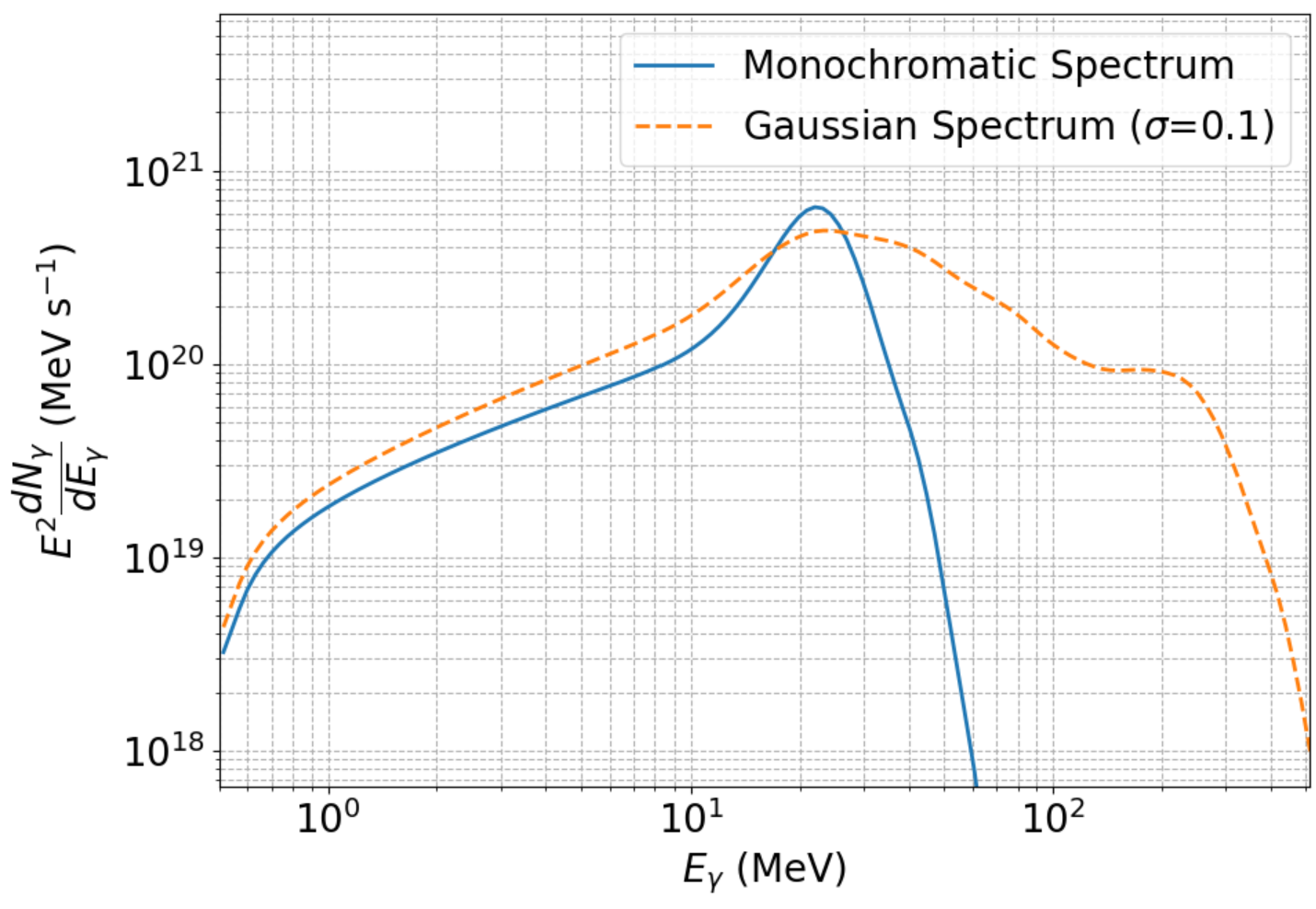}
    \caption{The total gamma-ray spectrum per PBH, from a PBH of mass $3 \times 10^{15}$ grams (blue line) and from a Gaussian distribution of density perturbations leading to a distribution of a mean mass of $3 \times 10^{15}$ grams. $\sigma$ refers to the standard deviation of initial density perturbations \cite{Biagetti:2021eep}.}
    \label{fig:PBHmass_distr_spectra}
\end{figure}

Users can calculate the gamma-ray spectra from four types of PBH mass distributions. Those are, i) a monochromatic distribution with a mass to be set in the range of $5\times 10^{13}$ to $1\times 10^{19}$ grams, ii) a Gaussian distribution of PBH masses originating from a Gaussian distribution of density perturbations \cite{Biagetti:2021eep}, iii) a more realistic non-Gaussian PBH mass distribution from \cite{Biagetti:2021eep} and iv) a log-normal distribution of PBH masses. 
In Fig.~\ref{fig:PBHmass_distr_spectra}, we give the gamma-ray spectra from monochromatic and Gaussian PBH mass-distributions.

\subsection{Software content}
\texttt{GammaPBHPlotter} was written in \texttt{Python} version 3.9 and is capable of running on Windows, Linux, and Mac. The main code uses five modules in its routine. Those being \texttt{colorama} \cite{hartley_colorama_046_2022}, \texttt{numpy} \cite{harris_array_numpy_2020}, \texttt{matplotlib} \cite{hunter_matplotlib_2007}, \texttt{tqdm}, \cite{dacostaluis_tqdm_zenodo_2024} and \texttt{scipy} \cite{virtanen_scipy_2020}. Since the software automatically checks and downloads all missing modules, this requirement should not be a concern for the user. We provide the software in \cite{ZenodoLink} that include the code an a relevant manual.

We acknowledge the use of \texttt{BlackHawk} \cite{Arbey:2021mbl}. This material is based upon work supported by the U.S. Department of Energy, Office of Science, 
Office of High Energy Physics, under Award No. DE-SC0022352.

\bibliography{sample}

\end{document}